# Lattice Boltzmann simulation of viscous fingering of immiscible displacement in a channel using an improved wetting scheme

Yuan Yu,[1,3,4] Haihu Liu,[2, a)] Yonghao Zhang,[3] and Dong Liang[1,4,b)]

[1] *School of Engineering, Sun Yat-sen University, Guangzhou 510006, China*

[2] *School of Energy and Power Engineering, Xi'an Jiaotong University, 28 West Xianning Road, Xi'an 710049, China*

[3] *Department of Mechanical and Aerospace Engineering, University of Strathclyde, Glasgow G1 1XJ, United Kingdom*

[4] *Guangdong Provincial Key Laboratory of Fire Science and Technology, Guangzhou 510006, China*

An improved wetting boundary implementation strategy is proposed based on lattice Boltzmann color-gradient model in this paper. In this strategy, an extra interface force condition is demonstrated based on the diffuse interface assumption and is employed in contact line region. It has been validated by three benchmark problems: static droplet wetting on a flat surface and a curved surface, and dynamic capillary filling. Good performances are shown in all three cases. Relied on the strict validation to our scheme, the viscous fingering phenomenon of immiscible fluids displacement in a two-dimensional channel has been restudied in this paper. High viscosity ratio, wide range contact angle, accurate moving contact line and mutual independence between surface tension and viscosity are the obvious advantages of our model. We find the linear relationship between the contact angle and displacement velocity or variation of finger length. When the viscosity ratio is smaller than 20, the displacement velocity is increasing with increasing viscosity ratio and reducing capillary number, and when the viscosity ratio is larger than 20, the displacement velocity tends to a specific constant. A similar conclusion is obtained on the variation of finger length.

## I. INTRODUCTION

Wetting phenomena are ubiquitous in natural and in industry. Numerical studies of phase interface are always based on diffuse interface model [1][2][3][4] or sharp interface model. It is known that [3][4] the diffuse interface model has obvious advantages when describing near-critical interfacial phenomena, dynamic contact line, contact angle hysteresis, small-scale flows, breakup or coalescence of droplet or bubble. On basis of phase interface dynamics, there are two ideas for imposing the appropriate wetting condition. One is the surface-energy formulation [1][5][6] imposed contact angle condition based on Young's equation of constant surface tension. This idea has been introduced into various multiphase lattice Boltzmann

---

[a)] Author to whom correspondence should be addressed. Electronic mail (a): haihu.liu@mail.xjtu.edu.cn; (b): gzliangd@163.com

communities [7][8][9]. The other idea is proposed by Ding et al. [3], the so-called geometric formulation for a prescribed contact angle $\theta_s$, which has been developed in lattice Boltzmann color-gradient model by Leclaire et al. [10] and Liu et al. [11].

The resulting contact angles of surface-energy formulation scheme different from the prescribed contact angle is observed by Qian et al. [12] and Ding et al.[3], and the non-physical mass transfer (NPMT) effect, which is defined as a fictive mass transfer beyond the theoretical phase interface, is very obvious especially when the prescribed contact angle is less than 90 degrees [9][10]. Physically, the precursor film is a beyond hydrodynamic scale effect, and we should not always observe the spreading beyond the nominal contact line region in the results of the surface-energy formulation. The root cause of evident NPMT effect in surface-energy formulation scheme is the conflict between two assumptions. In surface-energy formulation, some kind of fluid property $\varphi_s(x)$, e.g. density, is imposed on the virtual fluid nodes, then the surface tension between fluid and solid can be expressed. Substituting the surface tensions into Young's equation, we obtain a relationship between the fluid property $\varphi_s(x)$ and contact angle $\theta$. Noting that the Young's equation being based on the sharp interface assumption and working only in contact line region, if we imposed the virtual fluid property $\varphi_s(x)$ on all the solid surface nodes in a diffuse interface model, these two assumptions, i.e. sharp interface and diffuse interface, conflict each other. To resolve this conflict, we choose the geometric formulation for our diffuse interface multiphase model in this paper.

In the geometric formulation, Ding et al. [3][13] choose the tangential component $\boldsymbol{\tau} \cdot \boldsymbol{\nabla}\rho^N$ as the base, in other words, they introduce contact angle by only adjusting the normal component $\boldsymbol{n} \cdot \boldsymbol{\nabla}\rho^N = -|\boldsymbol{\tau} \cdot \boldsymbol{\nabla}\rho^N|\cot\theta$ at each time step. As thus, the NPMT effect no longer appears in single-phase region, but the base $|\boldsymbol{\tau} \cdot \boldsymbol{\nabla}\rho^N|$ is under the severe restriction of the isotropic order of DnQb lattice stencil especially in contact line region. When the contact angle is very small or very large, the base $|\boldsymbol{\tau} \cdot \boldsymbol{\nabla}\rho^N|$ and isotropic truncation error have the same order of magnitude, and which leads to failure. Improving the isotropic order or adopting eccentric isotropic difference can ease this deviation incompletely. To promote this geometric formulation into curved boundaries, Leclaire et al. [10] choose the modulus $|\boldsymbol{\nabla}\rho^N|$ as the base instead of the tangential component $|\boldsymbol{\tau} \cdot \boldsymbol{\nabla}\rho^N|$, then an unexpected improvement appeared, the base $|\boldsymbol{\nabla}\rho^N|$ is always large enough for the isotropic truncation error even in small or large contact angle cases. To estimate the unit normal vector $\boldsymbol{n}_f$ of $\boldsymbol{\nabla}\rho^N$ on the outermost fluid nodes, a linear least squares method is adopted in Leclaire's method, which works well in static contact angle problems, but it fails in dynamic problems because the vector $\boldsymbol{n}_f$ dramatically changes in contact line region, the linear or bilinear interpolation cannot use to fit the spatial function of vector $\boldsymbol{n}_f$. Up to now, there are still many places worth promoting. Firstly, we usually impose the contact angle boundary condition on the outermost fluid nodes, which is not the real physical position in both half-way bounce back scheme and full-way bounce back scheme. In a flat surface, we can use an extrapolation to ensure the wetting boundary condition is imposed on the real physical position [14][11], and lower spurious velocity results are obtained, but it's difficult to extend this extrapolation into complex geometric problems. Secondly, the mechanism of NPMT effect in contact line region is needed to be explored, which profoundly affect the stability and accuracy.

In this paper, an extra zero-interface-force condition is demonstrated based on the diffuse interface assumption in contact line region and is employed as an extra boundary condition. An improved wetting boundary implementation strategy is proposed based on lattice Boltzmann color-gradient model. This scheme is firstly validated by two static problems and one dynamic problem: a droplet resting on flat surface or a cylindrical surface and capillary filling in a two-dimensional (2D) channel. Then it is used to restudy the classical example, viscous fingering of immiscible fluids in a 2D channel, and new conclusions are made.

## II. NUMERICAL METHOD

### A. Lattice Boltzmann Multi-Relaxation-Time color-gradient model

A classic and effective color-gradient lattice Boltzmann model is adopted in this paper as developed in literature [15], which is developed on the basis of the works [16][17][18]. The present color-gradient model consists of three steps, i.e. the collision step, the recoloring step and the streaming step. Two distribution functions $f_{\alpha,R}$ and $f_{\alpha,B}$ are introduced to represent two immiscible fluids, i.e. red fluid and blue fluid. The total distribution function $f_\alpha = f_{\alpha,R} + f_{\alpha,B}$ undergoes a collision step as

$$f'_\alpha(x,t) = f_\alpha(x,t) + \Omega_\alpha(x,t) + \bar{F}_\alpha(x,t), \tag{1}$$

where $x$ and $t$ are the position and time respectively, $f'_\alpha$ is the post-collision distribution function, $\Omega_\alpha$ is the collision operator, $\bar{F}_\alpha$ is the forcing term, the subscript $\alpha$ means $\alpha$th direction of lattice velocity. In the MRT framework, the collision operator is given by

$$\Omega_\alpha(x,t) = -(M^{-1}SM)_{\alpha\beta}\left[f_\beta(x,t) - f_\beta^{eq}(x,t)\right], \tag{2}$$

where the subscript $\beta$ is the second dimension of second-order tensor, $M$ is a transformation matrix and is explicitly given [19] as

$$M = \begin{bmatrix} 1 & 1 & 1 & 1 & 1 & 1 & 1 & 1 & 1 \\ -4 & -1 & -1 & -1 & -1 & 2 & 2 & 2 & 2 \\ 4 & -2 & -2 & -2 & -2 & 1 & 1 & 1 & 1 \\ 0 & 1 & 0 & -1 & 0 & 1 & -1 & -1 & 1 \\ 0 & -2 & 0 & 2 & 0 & 1 & -1 & -1 & 1 \\ 0 & 0 & 1 & 0 & -1 & 1 & 1 & -1 & -1 \\ 0 & 0 & -2 & 0 & 2 & 1 & 1 & -1 & -1 \\ 0 & 1 & -1 & 1 & -1 & 0 & 0 & 0 & 0 \\ 0 & 0 & 0 & 0 & 0 & 1 & 1 & 1 & -1 \end{bmatrix}, \qquad (3)$$

and $S$ is a diagonal relaxation matrix and is given as

$$S = diag\left(\frac{1}{\tau_\rho}, \frac{1}{\tau_e}, \frac{1}{\tau_\varepsilon}, \frac{1}{\tau_j}, \frac{1}{\tau_q}, \frac{1}{\tau_j}, \frac{1}{\tau_q}, \frac{1}{\tau_v}, \frac{1}{\tau_v}\right), \qquad (4)$$

where $\tau_\rho$ and $\tau_j$ are related to the mass and momentum conserved equations respectively, we can choose any value for them without effect. $\tau_e$ and $\tau_\varepsilon$ are related to the internal energy, so they are unimportant for an isothermal or nearly isothermal problems. $\tau_v$ is given by

$$\tau_v(x,t) = \frac{v(x,t)}{c_s^2 \delta_t} + 0.5, \qquad (5)$$

where $v$ is the dynamic viscosity of the local fluid. In this paper, we choose, if there is no special instruction, the relaxation matrix for half-way bounce back scheme as

$$S = diag\left(1.0, 1.64, 1.54, 1.0, 1.9, 1.0, 1.9, \frac{1}{\tau_v}, \frac{1}{\tau_v}\right). \qquad (6)$$

It is worth noting that, the relaxation matrix should include the restriction $\frac{1}{\tau_q} = \frac{16\tau_v - 8}{8\tau_v - 1}$ to obtain exact non-slip condition in full way bounce back scheme, but there is no such restriction in half-way bounce back scheme [20][21], which is adopted in all cases in this paper.

The equilibrium distribution function $f_\alpha^{eq}$ in Eq. (2) is given by

$$f_\alpha^{eq}(\rho, \boldsymbol{u}) = \rho w_\alpha \left[1 + \frac{\boldsymbol{e}_\alpha \cdot \boldsymbol{u}}{c_s^2} + \frac{(\boldsymbol{e}_\alpha \cdot \boldsymbol{u})^2}{2c_s^4} - \frac{\boldsymbol{u}^2}{2c_s^2}\right], \qquad (7)$$

where $\rho = \rho_R + \rho_B$ is the total density with $\rho_R$ and $\rho_B$ being the densities of red and blue fluids, respectively; $c_s$ is the speed of sound; $\boldsymbol{e}_\alpha$ is the lattice velocity in the $\alpha$th direction, and $w_\alpha$ is the weight factor. For a 2D nine-velocity (D2Q9) model, $\boldsymbol{e}_\alpha$ is defined as $\boldsymbol{e}_0 = (0,0), \boldsymbol{e}_{1,3} = (\pm 1, 0), \boldsymbol{e}_{2,4} = (0, \pm 1), \boldsymbol{e}_{5,6} = (\pm 1, 1), \boldsymbol{e}_{7,8} = (\mp 1, -1)$; the speed of sound is defined as $c_s = \frac{\delta_t}{\sqrt{3}\delta_x}$ in D2Q9 model with $\delta_t$ and $\delta_x$ being the time step and space step respectively; the weight factor is given as $w_0 = \frac{4}{9}, w_{1,2,3,4} = \frac{1}{9}, w_{5,6,7,8} = \frac{1}{36}$.

The forcing term $\bar{F}_i$ in Eq. (1) contributes to the mixed interfacial regions and generates an interfacial tension. In the MRT framework, the forcing term is given by

$$\bar{F} = M^{-1}(I - 0.5S)M\tilde{F}, \tag{8}$$

where $I$ is a 9×9 unit matrix, $\bar{F} = [\bar{F}_0, \bar{F}_1, \bar{F}_2, \cdots, \bar{F}_3]^T$, and $\tilde{F} = [\tilde{F}_0, \tilde{F}_1, \tilde{F}_2, \cdots, \tilde{F}_8]^T$ is followed Guo's scheme [22] and is given by

$$\tilde{F}_\alpha = w_\alpha \left[\frac{e_\alpha - u}{c_s^2} + \frac{(e_\alpha \cdot u)e_\alpha}{c_s^4}\right] \cdot F\delta_t, \tag{9}$$

where the interfacial force $F$ is introduced by the perturbation operator as a body force based on the continuum surface force (CSF) model [23], which is expressed as

$$F = \frac{1}{2}\sigma\kappa\nabla\rho^N, \tag{10}$$

where $\sigma$ is surface tension coefficient, $\kappa$ is the local phase interface curvature, which is given by

$$\kappa = -\nabla_s \cdot n, \tag{11}$$

where $\nabla_s = (I - nn) \cdot \nabla$ is the surface gradient operator, $n = \frac{\nabla\rho^N}{|\nabla\rho^N|}$ is the unit normal vector of phase interface pointing into the red fluid, and the color gradient indicator function is defined by,

$$\rho^N(x,t) = \frac{\rho_R(x,t) - \rho_B(x,t)}{\rho_R(x,t) + \rho_B(x,t)}. \tag{12}$$

By simply reduction, the local interface curvature in 2D can be written as

$$\kappa = -n_x^2 \partial_y n_y - n_y^2 \partial_x n_x + n_x n_y (\partial_y n_x + \partial_x n_y). \tag{13}$$

The recoloring algorithm proposed by Latva-Kokko and Rothman [7] is also applied after the collision step and before the streaming step to promote phase segregation and maintain the interface. The recolored distribution functions of red and blue fluids, i.e. $f''_{\alpha,R}$ and $f''_{\alpha,B}$, are given as

$$\begin{aligned} f''_{\alpha,R}(x,t) &= \frac{\rho_R}{\rho} f'_\alpha(x,t) + \beta \frac{\rho_R \rho_B}{\rho} w_i \frac{e_\alpha \cdot \nabla\rho^N}{|\nabla\rho^N|}, \\ f''_{\alpha,B}(x,t) &= \frac{\rho_B}{\rho} f'_\alpha(x,t) + \beta \frac{\rho_R \rho_B}{\rho} w_\alpha \frac{e_\alpha \cdot \nabla\rho^N}{|\nabla\rho^N|}, \end{aligned} \tag{14}$$

where $\beta$ is a segregation parameter related to the interface thickness and is set to be 0.7 for numerical stability and model accuracy [24].

After the recoloring step, the streaming step for both red and blue distribution functions is executed as

$$f_{\alpha,k}(\mathbf{x} + \mathbf{e}_\alpha \delta_t, t + \delta_t) = f''_{\alpha,B}(\mathbf{x}, t), \quad k = R \text{ or } B. \tag{15}$$

With the post-streaming distribution functions, the density of each fluid is calculated by

$$\rho_k = \sum_\alpha f_{\alpha,k}, \quad k = R \text{ or } B, \tag{16}$$

and following Guo's forcing scheme [22], the other macroscopic physic quantity, the local fluid velocity is calculated by

$$\rho \mathbf{u}(x,t) = \sum_\alpha f_\alpha(\mathbf{x},t)\mathbf{e}_\alpha + \frac{1}{2}\mathbf{F}(\mathbf{x},t)\delta_t, \tag{17}$$

In this work, the red and blue fluids are assumed to have the same density for the sake of simplicity. To account for unequal viscosities of both fluids in the interface region continuously, a harmonic mean value [25][24] is adopted as

$$\frac{1}{\mu(\rho^N)} = \frac{1+\rho^N}{2\mu_R} + \frac{1-\rho^N}{2\mu_B}, \tag{18}$$

where $\mu_R$ and $\mu_B$ are the dynamic viscosities of red and blue fluids respectively.

**B. Interface force condition in contact line region**

In this part, a zero-interface-force condition in contact line region is demonstrated based on the diffuse interface model. This conclusion is then used as an extra Dirichlet condition in our implementation strategy. As shown in Figure 1, we consider an immiscible two-phase fluid adhering to the solid surface. The normal vector of solid surface is $\mathbf{n}_s$, the color gradient function is $\nabla \rho^N$ pointing into the red fluid side. The angle between vector $\mathbf{n}_s$ and vector $\nabla \rho^N$ is equal to the contact angle $\theta$, and the angle between vector $\nabla \rho^N$ and y-axis negative direction's unit vector $-\mathbf{j}$ is defined as $\zeta$, where $\zeta \in [0°, 180°]$. Giving the same conditions, i.e. the unit normal vector of solid surface $\mathbf{n}_s$ and the contact angle $\theta$, there are two states meet them. To strictly define the direction of vector $\nabla \rho^N$, we redefine the intersection angle as follows: if vector $\nabla \rho^N$ is in the clockwise direction of vector $-\mathbf{j}$, then use $\zeta_+$ to describe it; otherwise, we use $\zeta_-$. In the contact line region of diffuse interface model, we have

$$\partial_y \rho^N = f(\zeta) \partial_x \rho^N, \tag{19}$$

where

$$f(\zeta) = \begin{cases} \cot \zeta, & \zeta = \zeta_+ \\ -\cot \zeta, & \zeta = \zeta_- \end{cases}. \tag{20}$$

The direction of the intersection angle $\zeta$, i.e. clockwise $\zeta_+$ or counter-clockwise $\zeta_-$, decides the positive or negative of $\partial_x \rho^N$,

$$\begin{cases} \partial_x \rho^N < 0, & \zeta = \zeta_+ \\ \partial_x \rho^N > 0, & \zeta = \zeta_- \end{cases}. \tag{21}$$

As mentioned before, the unit normal vector of two fluids interface is given as

$$\boldsymbol{n} = \frac{\nabla \rho^N}{|\nabla \rho^N|} = P(x,y)\partial_x \rho^N \boldsymbol{i} + P(x,y)\partial_y \rho^N \boldsymbol{j} = n_x \boldsymbol{i} + n_y \boldsymbol{j}, \quad (22)$$

where $P(x,y) = \frac{1}{\sqrt{(\partial_x \rho^N)^2 + (\partial_y \rho^N)^2}}$.

Considering the first state $\zeta = \zeta_+$ in contact line region, we have

$$P(x,y) = -\frac{1}{\partial_x \rho^N} \frac{1}{\sqrt{1+(\cot \zeta)^2}} = -\frac{\sin \zeta}{\partial_x \rho^N}, \quad (23)$$

so the gradient of $P(x,y)$ can be written as

$$\begin{aligned} \partial_x P &= \frac{\sin \zeta}{(\partial_x \rho^N)^2} \partial_x^2 \rho^N \\ \partial_y P &= \frac{\sin \zeta}{(\partial_x \rho^N)^2} \partial_y \partial_x \rho^N \end{aligned}. \quad (24)$$

Combine Eq. (24) with Eq. (22), we have

$$\begin{aligned} \partial_x n_x &= -\frac{\sin \zeta}{\partial_x \rho^N} \cdot \partial_x^2 \rho^N + \frac{\sin \zeta}{(\partial_x \rho^N)^2} \partial_x^2 \rho^N \cdot \partial_x \rho^N = 0 \\ \partial_y n_x &= -\frac{\sin \zeta}{\partial_x \rho^N} \cdot \partial_y \partial_x \rho^N + \frac{\sin \zeta}{(\partial_x \rho^N)^2} \partial_y \partial_x \rho^N \cdot \partial_x \rho^N = 0 \\ \partial_x n_y &= -\frac{\sin \zeta}{\partial_x \rho^N} \cdot \partial_x(\partial_x \rho^N \cot \zeta) + \frac{\sin \zeta}{(\partial_x \rho^N)^2} \partial_x^2 \rho^N \cdot (\partial_x \rho^N \cot \zeta) = 0 \\ \partial_y n_y &= -\frac{\sin \zeta}{\partial_x \rho^N} \cdot \partial_x^2((\cot \zeta)^2 \rho^N) + \frac{\sin \zeta}{(\partial_x \rho^N)^2} \partial_x(\partial_x \rho^N \cot \zeta) \cdot (\partial_x \rho^N \cot \zeta) = 0 \end{aligned}. \quad (25)$$

Then considering the other state $\zeta = \zeta_-$ with the similar operations, we also obtain

$$\partial_x n_x = 0, \partial_y n_x = 0, \partial_x n_y = 0, \partial_y n_y = 0. \quad (26)$$

Combine Eq. (25), Eq. (26) with Eq. (10), we can obtain

$$\boldsymbol{F}(x) = \boldsymbol{0} \quad x \ in \ contact \ line \ region. \quad (27)$$

So we can make a conclusion now, based on the diffuse interface model and continuum surface force model, we have a zero-interface-force condition in the contact line region. It's an extra boundary condition which will be used in our wetting boundary implementation strategy later.

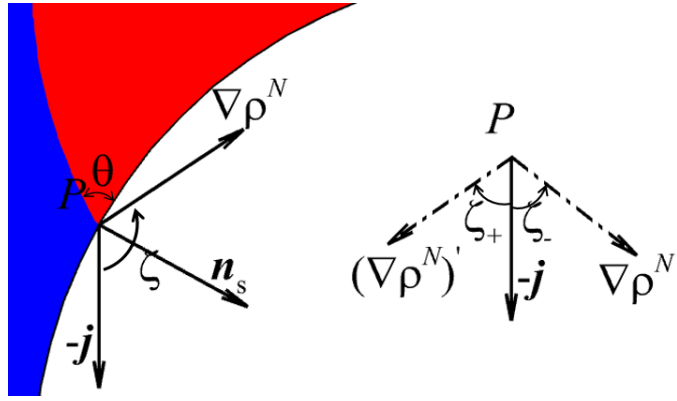

FIG. 1. Illustration for the angle relationship in mixed phase region.

### C. Generalized Wetting Boundary Implementation Strategy

In this section, we proposed a generalized boundary implementation strategy, which is applicable for complex boundary condition. There are two kinds of non-local operations in the color-gradient model, namely gradient $\nabla \rho^N$ and gradient $\nabla n_i$, where the subscript $i = x \text{ or } y$ in 2D problems as Eq. (22). On those fluid nodes which not contact with solid nodes, we use a fourth-order isotropic central difference to calculate gradient which is given by

$$\partial_i \varphi(x) = \frac{1}{c_s^2} \sum_\alpha w_\alpha \varphi(x + e_\alpha) e_\alpha, \tag{28}$$

where the field quantity $\varphi(x)$ can refer to $\rho^N$, $n_x$, and $n_y$. However, on the outermost fluid nodes, Eq. (28) is inappropriate since some values on solid nodes being needed, and there are two boundary conditions need to be imposed, i.e. $\frac{\nabla \rho^N}{|\nabla \rho^N|} = n_f$ and $\nabla n_i = 0$, where $n_f$ is the unit normal vector of interface on the outermost fluid nodes.

#### *1. The first boundary condition*

Considering the first boundary condition $\frac{\nabla \rho^N}{|\nabla \rho^N|} = n_f$ firstly, we consider the moduli $|\nabla \rho^N|$ and directions $n_f$ of gradients $\nabla \rho^N$ on the outermost fluid nodes individually. The gradients $\nabla \rho^N$ on the outermost fluid nodes can be approximately calculated by Eq. (30) if introducing a virtual fluid property $\rho^N$ on outermost solid nodes, which is defined by

$$\varphi_s(x) = \frac{\sum_{\alpha_1} w_{\alpha_1} \varphi(x + e_{\alpha_1})}{\sum_{\alpha_1} w_{\alpha_1}}, \tag{29}$$

where the subscript $\alpha_1$ is the $\alpha_1$th lattice velocity direction pointing into fluid nodes, and the subscript $s$ means on the outermost solid nodes. We can keep the moduli $|\nabla \rho^N|$ and obtain an estimated direction, which is written as

$$\left(n_f\right)_{es} = \frac{\nabla \rho^N}{|\nabla \rho^N|}, \tag{30}$$

where the subscript $es$ means an estimated direction which needs to be corrected later. Then we revise the directions by introducing the contact angle condition, and then the gradient $\nabla \rho^N$ on the outermost fluid nodes can be calculated by

$$\nabla \rho^N = |\nabla \rho^N| \boldsymbol{n}_f. \tag{31}$$

where $\boldsymbol{n}_f$ is the corrected direction of gradient $\nabla \rho^N$. Two steps are needed to obtain the correct direction $\boldsymbol{n}_f$ as shown below.

*a. Calculate the unit normal vector $\boldsymbol{n}_s$ of solid surface*

In curved boundary, the shape and the curvature of the solid surface is usually known, so the vector $\boldsymbol{n}_s$ is given without more processing. However, we don't know it in staircase boundary problems. As suggested by Xu et al.[26], an eighth-order isotropic discretization is introduced to estimate the vector $\boldsymbol{n}_s$, which is given by

$$\boldsymbol{n}_s(\boldsymbol{x}) = \frac{\sum_k w(|\boldsymbol{c}_k|^2) s(\boldsymbol{x} + \boldsymbol{c}_k \delta_t) \boldsymbol{c}_k}{|\sum_k w(|\boldsymbol{c}_k|^2) s(\boldsymbol{x} + \boldsymbol{c}_k \delta_t) \boldsymbol{c}_k|}, \boldsymbol{x} \in \Omega_{Fb}, \tag{32}$$

where $\boldsymbol{c}_k$ is the $k$th lattice velocity of eighth-order isotropic stencil as shown in Ref. [27], $s(\boldsymbol{x})$ is an indicator function which equals 0 for $\boldsymbol{x} \in \Omega_F$ and 1 for $\boldsymbol{x} \in \Omega_S$ and $w(|\boldsymbol{c}_k|^2)$ is the weight coefficient given by [27]

$$w(|\boldsymbol{c}_k|^2) = \begin{cases} \frac{4}{21} & |\boldsymbol{c}_k|^2 = 1 \\ \frac{4}{45} & |\boldsymbol{c}_k|^2 = 2 \\ \frac{1}{60} & |\boldsymbol{c}_k|^2 = 4, \\ \frac{2}{315} & |\boldsymbol{c}_k|^2 = 5 \\ \frac{1}{5040} & |\boldsymbol{c}_k|^2 = 8 \end{cases} \tag{33}$$

*b. Calculate the corrected unit normal vector $\boldsymbol{n}_f$ of fluid on physical solid surface*

To introduce the wetting contact angle $\theta$, a rotation matrix is used to obtain the prescribed direction of the phase interface normal vector. Considering the rotation direction, clockwise or counterclockwise as shown in Figure 1, vector $\boldsymbol{n}_f$ can be given by

$$(\boldsymbol{n}_f)_1 = \boldsymbol{n}_s \begin{bmatrix} \cos\theta & \sin\theta \\ -\sin\theta & \cos\theta \end{bmatrix}, (\boldsymbol{n}_f)_2 = \boldsymbol{n}_s \begin{bmatrix} \cos(-\theta) & \sin(-\theta) \\ -\sin(-\theta) & \cos(-\theta) \end{bmatrix}. \tag{34}$$

Then the Euclidean distances $D_1$ and $D_2$ are used to choose an appropriate theoretical direction, which are defined by

$$D_1 = \left| (\boldsymbol{n}_f)_1 - (\boldsymbol{n}_f)_{es} \right|, D_2 = \left| (\boldsymbol{n}_f)_2 - (\boldsymbol{n}_f)_{es} \right|, \tag{35}$$

And the unit vector of phase interface $\boldsymbol{n}_f$ is selected by

$$\boldsymbol{n}_f = \begin{cases} (\boldsymbol{n}_f)_1 & D_1 \leq D_2 \\ (\boldsymbol{n}_f)_2 & D_1 > D_2 \end{cases}, \tag{36}$$

### *2. The second boundary condition*

The second boundary condition $\nabla n_i = 0$ is the zero-interface-force condition demonstrated before. Introducing the virtual fluid property $n_i$ on outermost solid nodes by Eq. (29), we can calculate the $\nabla n_i$ on the outermost fluid nodes. This operation exactly implicitly imposed the condition $\nabla n_i = 0$, and a strict proof process is given below.

The weighted average virtual scheme given as Eq. (29) can be rewritten as

$$\sum_{\alpha_1} w_{\alpha_1} [\varphi(x + e_{\alpha_1}) - \varphi(x)] = 0. \tag{37}$$

Making a Taylor expansion, and cutting off the second order residual,

$$\sum_{\alpha_1} w_{\alpha_1} e_{\alpha_1 i} \partial_i \varphi(x) = 0, \tag{38}$$

Let,

$$\boldsymbol{a} = \sum_{\alpha_1} w_{\alpha_1} e_{\alpha_1 i}, \tag{39}$$

$$\boldsymbol{b} = \partial_i \varphi(x). \tag{40}$$

Obviously $\boldsymbol{a} \neq \boldsymbol{0}$, so

$$\boldsymbol{b} = \boldsymbol{0} \text{ or } \boldsymbol{b} \perp \boldsymbol{a}, \tag{41}$$

where $\boldsymbol{a}$ is deemed to be the normal direction of the liquid-solid surface, so $\boldsymbol{b} = \partial_i \varphi(x)$ always along the tangential direction of the liquid-solid surface when $\boldsymbol{b} \neq \boldsymbol{0}$. In either case, we have

$$\partial_n \varphi(x) = 0, \tag{42}$$

where $\boldsymbol{n}$ is the unit vector of the liquid-solid surface. Reconsidering the similar operation as section B, there is a linear relationship between the two components of gradient $\nabla \varphi(x)$, which can be given as

$$\partial_y \boldsymbol{n} = -f(\beta) \partial_x \boldsymbol{n}. \tag{43}$$

So we can deduce the following conclusion from Eq. (29),

$$\nabla n_i = 0, \tag{44}$$

### III. NUMERICAL SIMULATIONS

To measure the results of simulations, firstly, we defined a function $E_1$ to control the timing of iteration stop, which is given as

$$E_1 = max(|u_x^{t+500} - u_x^t|, |u_y^{t+500} - u_y^t|), \tag{45}$$

If $E_1 < 10^{-7}$, iteration is stopped, and we calculate the maximum spurious velocity in steady problems by

$$|\boldsymbol{u}|_{max} = max\left(\sqrt{u_x^2 + u_y^2}\right). \tag{46}$$

Besides, a NPMT function $E_{NPMT}$ is also defined in steady problems by

$$E_{NPMT} = \sqrt{\left(\frac{m_R^{inc}}{m_R^{all}}\right)^2 + \left(\frac{m_B^{outc}}{m_B^{all}}\right)^2}, \tag{47}$$

where $m_R$ and $m_B$ are the nominal mass of red fluid and blue fluid respectively, calculated by summation of $\rho_R$ or $\rho_B$ in a certain region, the superscript are the regions, $inc, outc, all$ mean summation in the region in the circle C, outside the circle C, and in all fluid nodes region respectively, where the circle C is a theoretical steady-state phase interface.

### A. Validation of a static droplet wetting on a flat surface

In this part, simulations are performed in a 160×100 lattice domain, and the initial radius of the droplet on the bottom wall is $R = 45$. For the sake of reducing the iteration steps and measuring the NPMT effect $E_{NPMT}$, the center of the droplet is initialized as $(x_c, y_c) = \left(\frac{160}{2}, \frac{3}{2} - R\cos\theta\right)$ in half-way bounce back scheme. To test the model effectiveness in kinematic viscosity ratio problems, several different kinematic viscosity ratios $\left(M = \frac{v_R}{v_B}\right)$ are investigated with $v_R = 0.35$. The surface tension coefficient is $\sigma = 0.02$. Considering the periodic boundary conditions can offset the NPMT effect to some extent[10], and in many practical applications, this offset does not exist, so we choose a closed boundary condition for all cases. Results of calculated contact angles are shown in Table I, and the errors are all very small.

TABLE I. Calculated contact angles in flat surface.

| M | 30° | 60° | 90° | 120° | 150° |
|---|---|---|---|---|---|
| 1 | 30.96 | 60.20 | 90.05 | 120.18 | 151.77 |
| 100 | 29.99 | 60.01 | 90.00 | 120.09 | 151.02 |

#### 1. Comparison between present and previous methods

In this part, we make a comparison among the present model, the surface-energy formulation model and the original geometric formulation model in flat surface problem. Different operations among these three models are in the part of imposing the wetting boundary conditions. In surface-energy formulation model, we set the virtual fluid property on the outermost solid nodes with $\rho^N = \cos\theta$ [14]. In original geometric formulation model, we consider the implementation in Ref. [3]. The viscosity ratio is fixed as $M = 1$. And the contact angle is set as $30°, 60°, 90°, 120°, 150°$. Comparison of the NPMT effects with iteration is made, results are shown in Figure 2. The present method (red lines) can effectively control the NPMT effect.

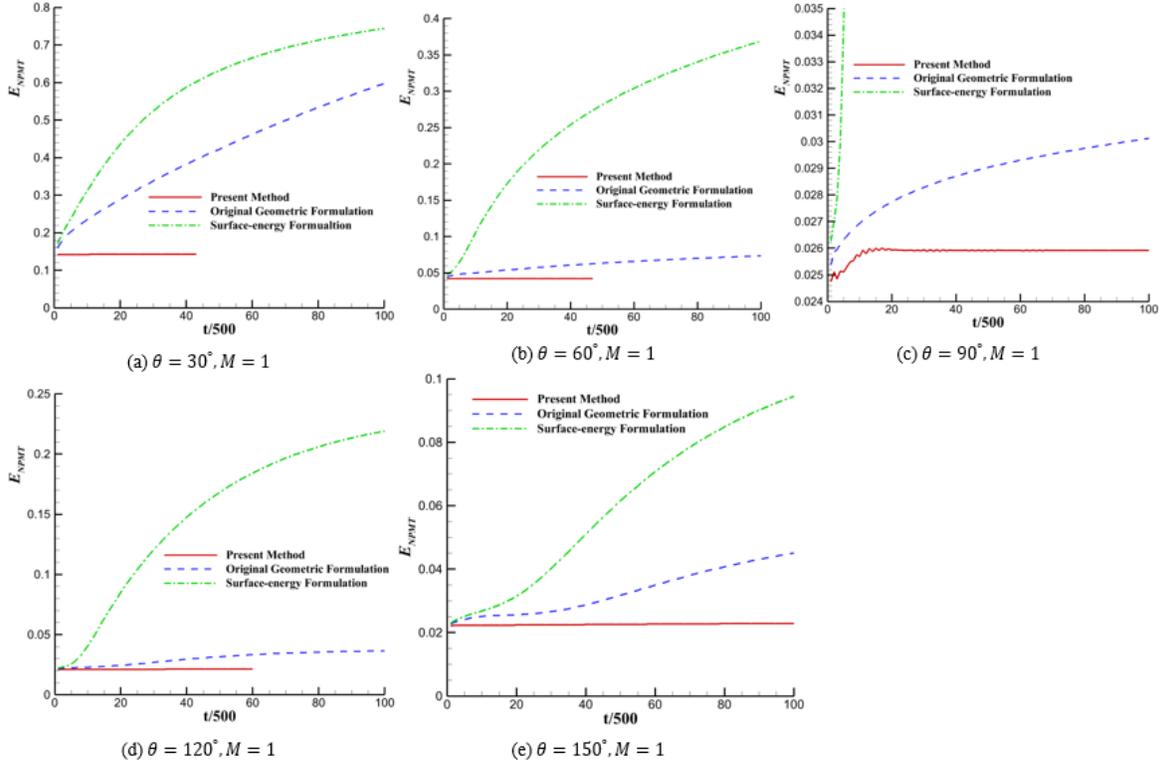

FIG. 2. The NPMT effect comparison among the present model (red solid line), the original geometric formulation model (blue dot line) and the surface-energy formulation model (green dash-dot line) with different contact angles.

### *2. Influence of zero-interface-force condition in contact line region*

The zero-interface-force condition in contact line region is firstly proposed and imposed on boundary as an extra condition, we will test its (called *scheme (a)*) influence by comparing to the *scheme (b)* which ignored this condition and calculated the gradient $\nabla n_i$ on the outermost fluid nodes by

$$\partial_i \varphi_b(\boldsymbol{x}) = \frac{\sum_{\alpha_1} w_{\alpha_1}[\varphi(\boldsymbol{x}+\boldsymbol{e}_{\alpha_1})-\varphi(\boldsymbol{x})]e_{\alpha i}}{\sum_{\alpha_1} w_{\alpha_1}}, \tag{48}$$

where the subscript $\alpha_1$ means $\alpha_1$th the lattice velocity direction pointing into fluid nodes. Different viscosity ratios are set as $M = 1, 100$, and Different contact angles, i.e. $30°, 60°, 90°, 120°, 150°$, are also tested. Results are shown in Table II, the extra condition $\nabla n_i = 0$ indeed reduced the spurious velocity.

TABLE II. Influence of zero-interface-force condition in contact line region.

| $M$ | $|\boldsymbol{u}|_{max}$ | 30° | 60° | 90° | 120° | 150° |
|---|---|---|---|---|---|---|
| 1 | *scheme (a)* | 1.79 | 0.51 | 0.18 | 0.59 | 1.47 |
|   | *scheme (b)* | 1.80 | 0.53 | 0.22 | 0.61 | 1.48 |
| 100 | *scheme (a)* | 33.4 | 11.9 | 1.86 | 6.29 | 10.3 |
|   | *scheme (b)* | 33.8 | 12.1 | 2.19 | 6.39 | 10.3 |

a$|\boldsymbol{u}|_{max} \times 10^{-4}$.

## B. Validation of a static droplet wetting on a cylindrical surface
### *1. Basic Validation*

To test the present model in curved boundary problems, we set a droplet resting on a cylindrical surface. Simulations are performed in a 200×200 lattice domain, the radius of cylinder and droplet are set as R = 40, the solid circle is located on (100, 60), the center of the droplet is located on $(100, 2R \sin \theta + 60)$, and the viscosity ratio is fixed as 1. Different prescribed contact angles, i.e. $10°, 30°, 45°, 60°, 90°, 120°, 135°, 150°, 170°$ are chosen. The stable state results are given in Figure 3, which are all fitting well with the theoretical solution shown as dash-dot circles. The maximum spurious velocities for all contact angles are shown in Table III, which are all in the order of magnitude $10^{-4}$. However, it is worth noting that, when the contact angle is larger than or equal to $150°$, the droplet slightly upward. It is because there are only several lattices between solid and red fluid contact region, which even less than the thickness of phase interface, and if densified grids, this error will be offset.

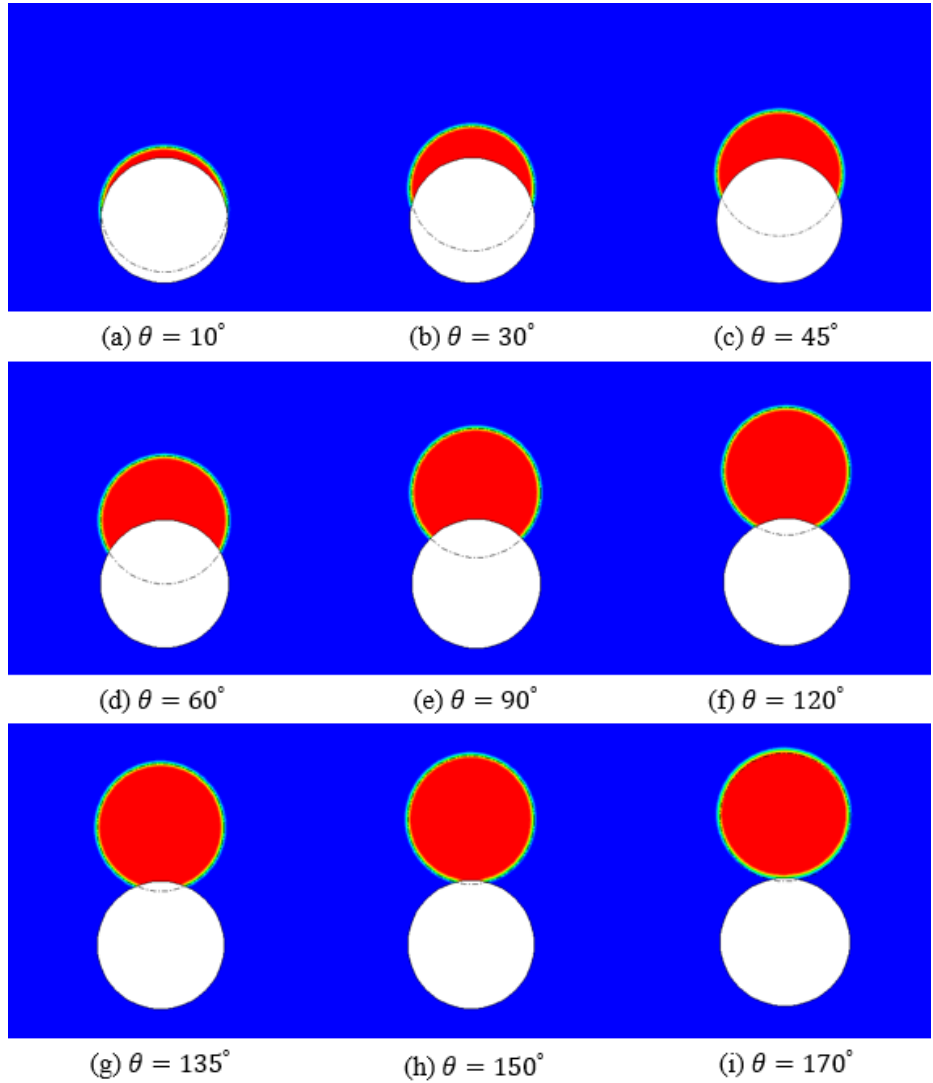

FIG. 3. A droplet resting on a cylindrical surface with different contact angles, i.e. $\mathbf{10°, 30°, 45°, 60°, 90°, 120°, 135°, 150°, 170°}$.

TABLE III. The maximum spurious velocity comparison between *scheme (a)* and *scheme (b)* to test the effect of zero-interface-force condition in contact line region (curved surface).

| $\|u\|_{max}(\times 10^{-5})$ | 10° | 30° | 45° | 60° | 90° | 120° | 135° | 150° | 170° |
|---|---|---|---|---|---|---|---|---|---|
| Scheme (a) | 1.51 | 1.19 | 1.40 | 2.26 | 6.10 | 16.5 | 22.4 | 30.6 | 26.5 |

### 2. Effect of the eighth-order isotropic discretization estimation scheme

In our model, it is recommended [26] to use an eighth-order isotropic discretization to estimate the unit normal vector $\boldsymbol{n}_s$ of solid surface. The problem of a droplet resting on cylindrical surface can be treated as a staircase surface with the eighth-order isotropic discretization estimation or a curved surface with the theoretical unit normal vector of solid. We make a comparison between scheme (a) staircase surface and scheme (b) curved surface in the cylindrical problem to test the effect of the eight-order isotropic discretization. Different contact angles, i.e. 30°, 60°, 90°, 120°, 150°, are tested. The other parameters are same as before. Results are given in Table IV: the NPMT effects of both treatments are close, the eighth-order isotropic discretization basically meet the need of simulation; the spurious velocity of staircase surface is larger than curved surface as the contact angle away 90°, which means the eight-order isotropic discretization is very effective as the contact angle near 90°, but not very good when the contact angle is very large or very small.

TABLE IV. Effect of the eight-order isotropic discretization estimation scheme.

| $\theta$ | $\|u\|_{max}(\times 10^{-4})$ | | $E_{NPMT}(\times 10^{-2})$ | |
|---|---|---|---|---|
| | staircase | curved | staircase | curved |
| 30° | 3.56 | 0.12 | 3.55 | 4.44 |
| 60° | 1.36 | 0.23 | 2.72 | 2.70 |
| 90° | 0.65 | 0.61 | 2.36 | 2.23 |
| 120° | 2.21 | 1.65 | 2.32 | 2.32 |
| 150° | 4.89 | 3.07 | 2.93 | 2.76 |

### C. Validation in dynamic problem: capillary filling

Capillary intrusion is a good benchmark example for assessing whether a multiphase model is able to simulate moving contact line problems [11], especially in our model, we impose a zero-interface-force condition in contact line region. Neglecting the gravity and inertial effects, the balance among the intruding fluid viscous drag, the Laplace pressure and the pressure difference over the interface can be written as [28]

$$\sigma \cos\theta = \frac{6}{d}[\mu_R x + \mu_B(L-x)]\frac{dx}{dt}, \qquad (49)$$

where $\theta$ is the contact angle and is set as 10°, 30°, 45°, 60° for the red fluid, regard as wetting fluid in this part, $d$ is the width of capillary tube and is set as 21 in half-way bounce back scheme, $\sigma$ is the surface tension coefficient and is set as 0.005, $x$ is the position of the moving interface with $x = 0$ at the inlet of capillary tube. $L$ is the length of capillary tube and is set as 200, and $\mu_R$ and $\mu_B$ are the dynamic viscosities of the red and blue fluids respectively. The whole simulation domain is in a 400×35 system with periodic boundary condition and half-way bounce back scheme is adopted on solid surface. Two different viscosity ratios: (a) $M = 1$ and (b) $M = 100$ are tested. The solid domain set at the position $100 \leq x \leq 300$ with thickness as 7, and in fluid domain, the blue fluid is initialized at the position $120 \leq x \leq 375$, and the remaining lattice sites are filled with red fluid as shown in Figure 4. Results are shown in Figure 5. All results match better with theory solution than using the method proposed Leclaire et al. [10].

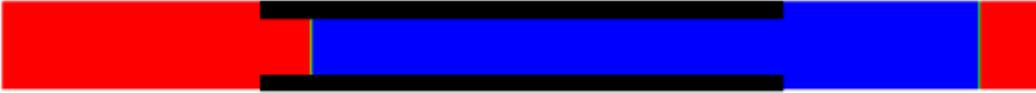

FIG. 4. Illustration of capillary intrusion initialization.

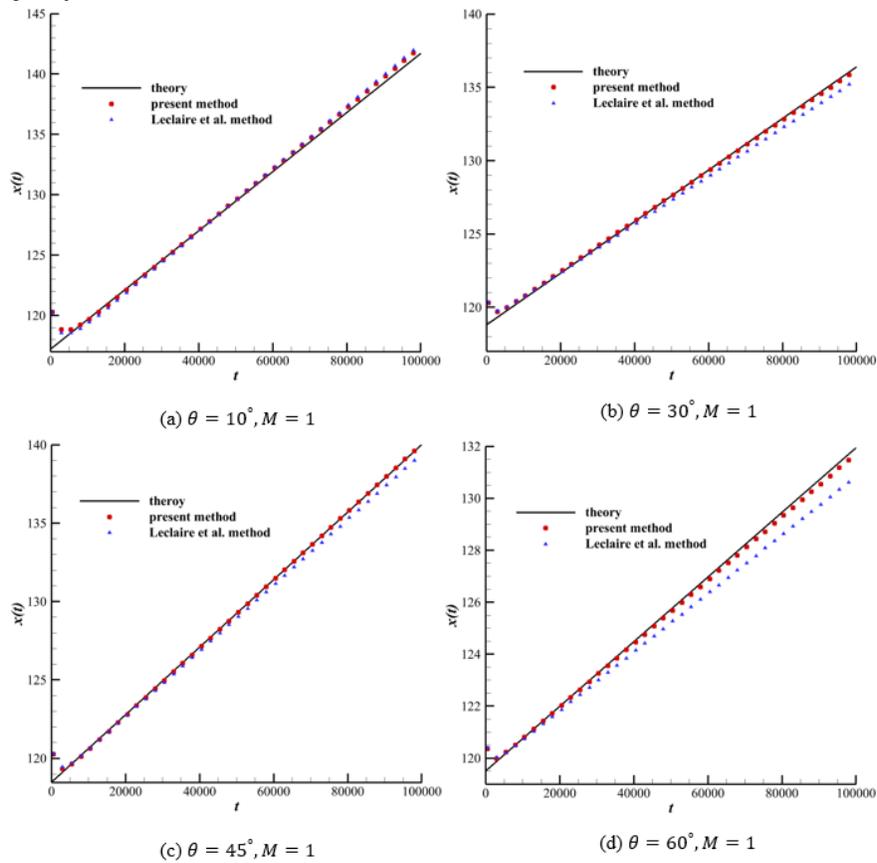

(a) $\theta = 10°, M = 1$

(b) $\theta = 30°, M = 1$

(c) $\theta = 45°, M = 1$

(d) $\theta = 60°, M = 1$

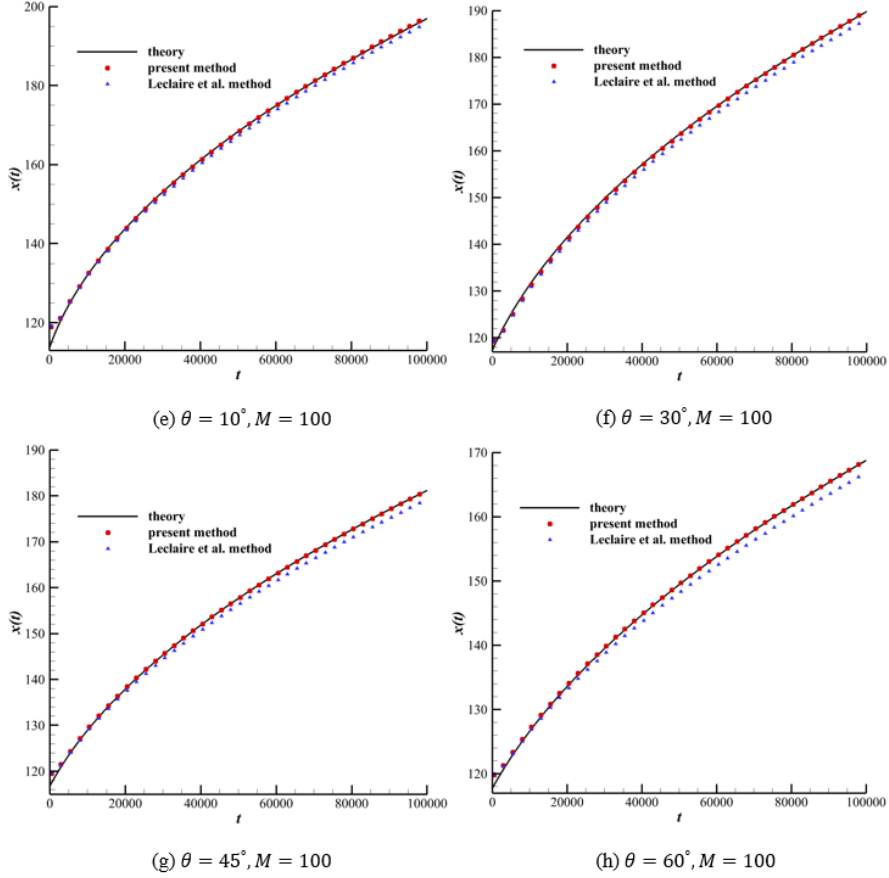

(e) $\theta = 10°, M = 100$  (f) $\theta = 30°, M = 100$

(g) $\theta = 45°, M = 100$  (h) $\theta = 60°, M = 100$

FIG. 5. The position of phase interface $x(t)$, versus iteration time for different contact angles and viscosity ratios. The (dark) solid lines are the theoretical predictions for Eq. (49), the (red) open circles represent the simulation results of the present method, and the (blue) open deltas represent the simulation results using the method proposed by Leclaire et al.[10].

**D. Viscous fingering of immiscible fluids displacement in a channel**

The viscous fingering phenomenon of immiscible fluids displacement in a 2D channel is a classic example before we study the displacement phenomenon in porous media. It has been similarly studied by Chin et al.[29], Kang et al. [30], Dong et al. [31] and Shi et al. [32] using different lattice Boltzmann model. In Ref.[29][30][31], the Shan-Chen multiphase lattice Boltzmann method is used to simulate this problem, but the surface tension is impossible to tune independently since it is coupling with the kinematic viscosity and density ratio in their scheme. In Ref. [32], the free energy lattice Boltzmann method is used and only 90° and 180° are considered, however, the wetting boundary condition has not been strictly verified, which impact clearly on the velocity of moving contact line. In this part, we systematically investigate the effects of capillary number, surface tension, kinematic viscosity, viscosity ratio, wettability and inlet velocity in this problem. In our model, all the parameters can be independently tunable without coupling, high viscosity ratios and wide range contact angles can be studied, and the speed of moving contact line is credible which has been validated before.

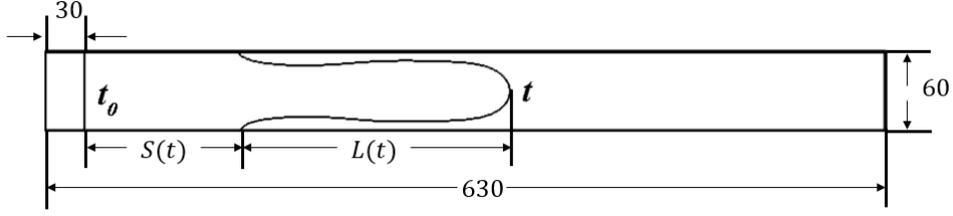

FIG. 6. Schematic illustration of simulation geometry.

In addition to the first comparison part, as shown in Figure 6, simulations are performed in a 630×60 domain, and the initial interface is setting on $x = 30$ to ensure the curved interface staying in computational domain as the contact angle being small. The displacing (red) fluid is on the left, and the displaced (blue) fluid is on the right. On the left inlet, a Poiseuille velocity profile with maximum value $u_0$ scheme proposed by Zou and He [33] is imposed, on the right outlet, the open boundary conditions are employed, and on the solid boundary, the halfway bounce back boundary scheme is applied to achieve the non-slip velocity condition [20]. Two parameters are used to measure the interface moving: $S(t)$ is the moving distance of contact line to the initial position $S(t_0) = 0$, and $L(t)$ is the finger length which is measured on the horizontal center axis of the channel.

### *1. Comparison with results of Kang et al.*

In this part, we set similar parameters to make a comparison with results of Kang et al. [30], which has been used to make a comparison by Dong et al. [31] and Shi et al. [32]. The same initialization conditions are set as: the kinematic viscosities are $v_R = 1/3$ and $v_B = 1/12$, with the same unity density being using for both displacing (red) and displaced (blue) fluid and the viscosity ratio is $M = 4$; a Poiseuille velocity profile with maximum value $u_0$ is enforced at both inlet and outlet to avoid the boundary effects and to develop a steady finger; the maximum driving Poiseuille velocity is chosen as $(a)$ $u_0 = 0.01$, $(b)$ $u_0 = 0.05$, and $(c)$ $u_0 = 0.1$; correspond to the choosing $(a)$ $\Delta t = 5000$, $(b)$ $\Delta t = 1000$, and $(c)$ $\Delta t = 500$; and the contact angle is $\theta = 90°$. The different settings are: a domain with grid size 410×66 different from 400×66 in Ref. [30] is used with the initial phase interface on $x = 10$ instead of $x = 0$ and an initial density ratio of 3% in Ref. [30]; the surface tension is fixed on $\sigma = 0.0462$ to match the capillary numbers in Ref. [30], and the surface tension in Kang's paper is not given, where the capillary number can be calculated by $Ca = \frac{v_R u_0}{\sigma}$ in this paper. Results are given in Figure 7. The morphology of present results matches well with results of Kang et al. [30].

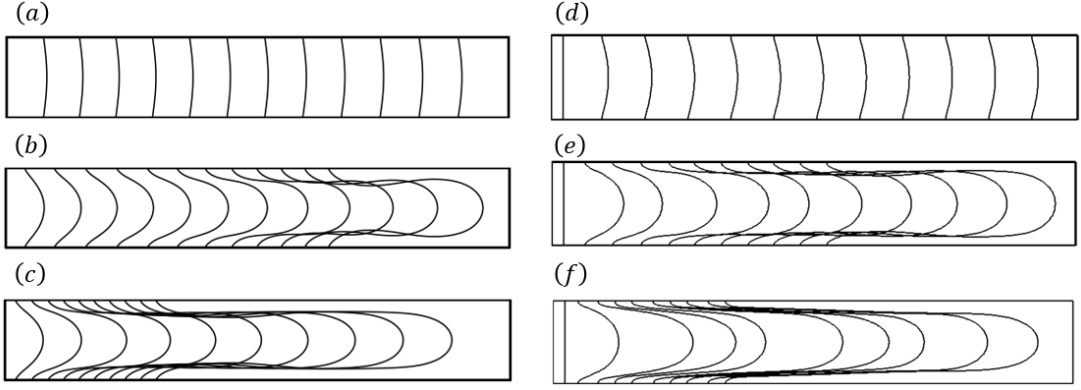

FIG. 7. The phase interface morphology comparison between results from present method (right) and results in Kang et al. [34] (left) with $(a), (d)$ $u_0 = 0.01, \Delta t = 5000$, $(b), (e)$ $u_0 = 0.05, \Delta t = 1000$, $(c), (f)$ $u_0 = 0.1, \Delta t = 500$, and with fixed parameters: $M = 1$, $\sigma = 0.0462$, $v_R = 1/3$, and $\theta = 90°$.

### 2. Effects of wettability

In this part, the effects of wettability, different contact angles, i.e. $20°\sim 160°$, are investigated. The other parameters are set as: the viscosities are $v_R = 0.3$ and $v_B = 0.3$ with the viscosity ratio $M = 1$; the maximum driving velocity is $u_0 = 0.1$; the surface tension is $\sigma = 0.06$ with capillary number $Ca = 0.5$, and the displacement distance $S(t)$ and the finger length $L(t)$ as shown in Figure 6. are used to measure displacement behavior. Results are shown in Figure 8. The displacement velocity and variation speed of finger length, i.e. $\frac{dS}{dt}$ and $\frac{dL}{dt}$ respectively, keep constant in all cases. There are evident linear functions between the variation speeds, i.e. $\frac{dS}{dt}$ and $\frac{dL}{dt}$, and contact angle, and they can be expressed as

$$\frac{dS}{dt}(\theta) = 5.278 \times 10^{-2} + 1.280 \times 10^{-4} \times \theta, \tag{50}$$

$$\frac{dL}{dt}(\theta) = 3.095 \times 10^{-2} - 1.212 \times 10^{-4} \times \theta, \tag{51}$$

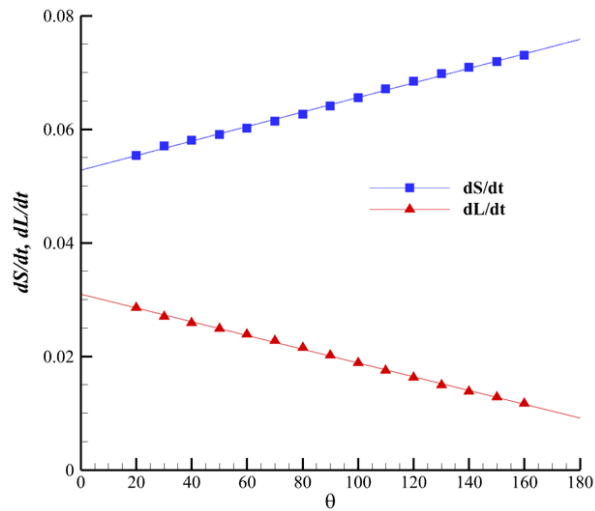

FIG. 8. The variation speeds in steady state of displacement distance and finger length, i.e. $dS/dt$ and $dL/dt$, as a function of the contact angle $\theta$ with $Ca = 0.5$, $M = 1$, $u_0 = 0.1$ and $\sigma = 0.06$. The slopes are $1.280 \times 10^{-4}$ and $-1.212 \times 10^{-4}$, respectively.

### 4. Effects of capillary number and viscosity ratio

In this part, the effects of capillary number $Ca$ and viscosity ratio $M$ are investigated. The capillary numbers $Ca$ are set as 0.72, 0.60, 0.48 and 0.36, and the viscosity ratios $M$ are set as 1, 2, 3, 5, 10, 20, 50, 100, 200, 500, 1000. The other parameters are set as: the maximum driving velocity is $u_0 = 0.01$; the viscosity of red fluid is $v_R = 0.5$; the contact angle is fixed at $\theta = 90°$. After the variation speeds reach steady state, we recorded variation speeds $\frac{dS}{dt}$ and $\frac{dL}{dt}$.

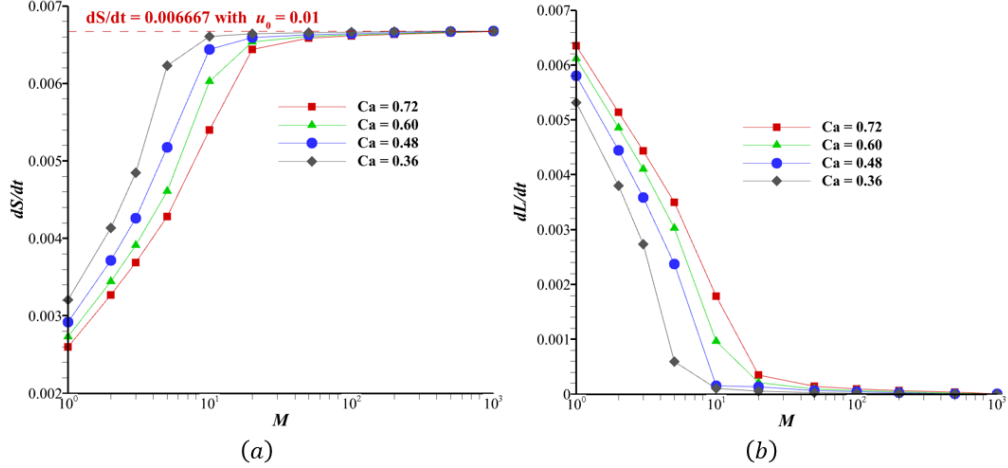

FIG. 9. Effects of capillary number and viscosity ratio with fixed parameter: $u_0 = 0.01$, $v_R = 0.5$, and $\theta = 90°$.

As shown in Figure 9., when the viscosity ration $M$ is smaller than 20, the displacement velocity (a) and variation speed of finger length (b) are changing monotonically, increasing and decreasing respectively, with increasing viscosity ratio; the displacement velocity is reducing with increasing capillary number, whereas the variation speed of finger length. When the viscosity ratio is larger than 20, the variation speeds of displacement and the finger length maintain constants. As shown in Figure 9(a), the speed of displacement tends to 0.00667 with increasing viscosity ratio under present parameters setting, which is exactly 1/3 of maximum driving velocity $u_0 = 0.01$. As shown in Figure 9(b), the finger length maintains a constant as viscosity ratio lager than 20, the equilibrium finger lengths $L_{eq}$ versus different viscosity ratios and capillary numbers are shown in Figure 10(a), a short finger length $L$ will arise as a small capillary number and a high viscosity ratio, by the way, the morphology of phase interface of high viscosity ratio, i.e. 1000, 500, 200, 100, 50, 20 from left to right, with $Ca = 0.72$ are shown in Figure 10(b).

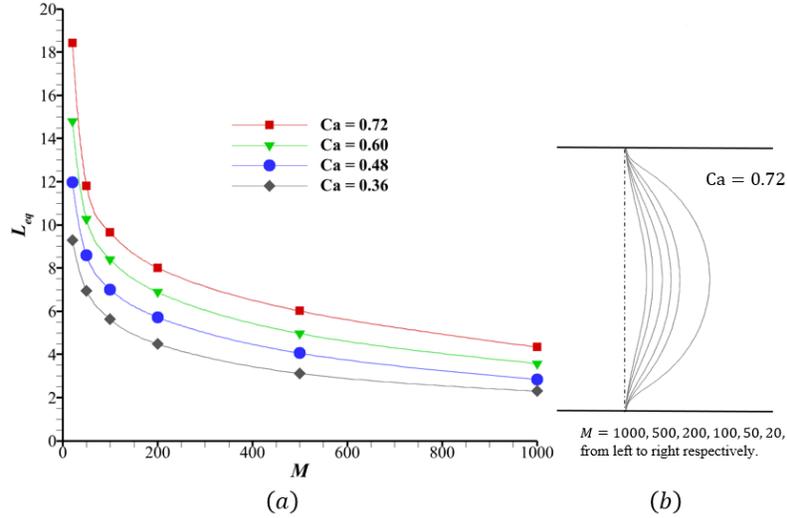

FIG. 10. The equilibrium finger length $L_{eq}$ versus different high viscosity ratios and capillary numbers.

**IV. CONCLUSION**

In this work, an improved wetting boundary condition implementation scheme is proposed based on the color-gradient lattice Boltzmann method of Liu et al. [11]. A zero-interface-force condition is derived based on the diffuse interface assumption in contact line region, and is employed as an extra boundary condition. This scheme is validated by two static problems and one dynamic problem: a droplet resting on flat surface or a cylindrical surface and capillary filling in a 2D channel. In static flat surface problems, the present scheme can control the non-physical mass transfer effect effectively, and smaller spurious velocity is obtained. In static curved surface problems, the present scheme can simulate all contact angels effectively, and the eighth-order isotropic discretization estimation method for the unit normal vector of solid surface proposed by Xu et al. [26] has been evaluated, good results can be obtained especially as the contact angle close to $90°$. In dynamic problems, our scheme also shows better performance than the method proposed by Leclaire et al. [10] and our results agree well with the theoretical solutions with different contact angles and viscosity ratios.

Based on the strict validations of the present scheme, we restudy the classical example, viscous fingering of immiscible fluids in a 2D channel. Several affecting factors, i.e. surface tension, viscosity, viscosity ration, inlet velocity, wettability, and capillary number, are tuning independently to investigate. The contact angle linearly affects the variation speed of displacement distance and the variation speed of finger length as viscosity ratio $M = 1$. The variation speed of displacement distance increasing with the increasing viscosity ratio and reducing capillary number when the viscosity ratio is smaller than 20, however, when the viscosity ratio is larger than 20, the displacement process tends to be a constant velocity, which is exactly two-third of the maximum Poiseuille velocity under our parameter settings. The variation speed of finger length is also showing similar rule, the finger length will maintain an equilibrium state as viscosity ratio larger than 20, and the equilibrium finger length reduces with the increasing viscosity ratio and reducing capillary number.

**ACKNOWLEDGMENTS**

This work was support by funds of This work was supported by the National Natural Science Foundation of China (No. 51506168), the National Key Research and Development Project of China (No. 2016YFB0200901), the China Postdoctoral Science Foundation (No. 2016M590943), and the HPCC platform of Xi'an Jiaotong University. H.L. gratefully acknowledges the financial supports from Thousand Youth Talents Program for Distinguished Young Scholars, the Young Talent Support Plan of Xi'an Jiaotong University, Guangdong Provincial Key Laboratory of Fire Science and Technology (No. 2010A060801010) and Guangdong Provincial Scientific and Technological Project (No. 2011B090400518).